\documentclass[twocolumn,superscriptaddress,showpacs,floatfix,preprintnumbers, nofootinbib,hyperref]{revtex4} 
\usepackage{graphicx}
\usepackage{rotating}
\usepackage{epsfig}
\usepackage{bm}
\usepackage[usenames,dvipsnames]{xcolor}

\usepackage{color}

\def\he4{$^4$He}
\def\h2{$^2$H}

\newcommand{\lesssim}{\,\rlap{\lower3.7pt\hbox{$\mathchar\sim$}}
\raise1pt\hbox{$<$}\,}

\begin{document}


\title{{ Multi-azimuthal-angle
instability  for different supernova neutrino fluxes}}

\author{Sovan Chakraborty}
\affiliation{Max-Planck-Institut f\"ur Physik
(Werner-Heisenberg-Institut)\\
 F\"ohringer Ring 6, D-80805 M\"unchen, Germany}

\author{Alessandro Mirizzi} 
\affiliation{II Institut f\"ur Theoretische Physik, Universit\"at Hamburg, Luruper Chaussee 149, 22761 Hamburg, Germany}


\begin{abstract}
It has been recently discovered that removing the axial symmetry in the  ``multi-angle effects'' associated
with the neutrino-neutrino interactions for supernova (SN) neutrinos, a new multi-azimuthal-angle (MAA) instability 
would trigger  flavor conversions  in addition to the ones caused by the bimodal and multi-zenith-angle (MZA) instabilities.
We investigate the dependence of the MAA instability on the original SN neutrino
fluxes, performing a  stability analysis
of the linearized neutrino equations of motion.
We compare these results with the numerical evolution
of the SN neutrino non-linear equations, 
looking at a local solution along a specific line of sight, 
under the assumption that the transverse variations of the global solution are small. We also assume 
 that self-induced conversions are not suppressed 
by large matter effects.  We show that the pattern of the spectral crossings (energies
where $F_{\nu_e} = F_{\nu_x}$, and  $F_{\bar\nu_e} = F_{\bar\nu_x}$) is crucial in determining the impact of MAA effects
on the flavor evolution. For neutrino spectra  with a strong excess of 
$\nu_e$ over $\bar\nu_e$, presenting only a single-crossing,  MAA instabilities would trigger new flavor conversions in normal mass hierarchy. In our simplified flavor evolution scheme, these would lead
to spectral swaps and splits analogous  to what produced in inverted hierarchy by the bimodal instability. 
Conversely, in the presence of spectra with a moderate flavor hierarchy, having  multiple crossing energies, MZA effects 
would produce a sizable delay in the onset of the
flavor conversions, inhibiting the growth of the MAA instability. In this case the splitting features for the oscillated spectra 
in both the mass hierarchies
are the ones induced by the only bimodal and  MZA effects.

\end{abstract}


\maketitle

\section{Introduction}

Supernova (SN) neutrinos represent  a subject of
large interest in astroparticle physics~\cite{Raffelt:2012kt}. In particular,
renewed attention is being paid to collective features of
flavor transformations~\cite{Qian:1994wh,Samuel:1993uw,  
  Kostelecky:1993dm, Kostelecky:1995dt, Samuel:1996ri, Pastor:2001iu,  
  Pastor:2002we, Sawyer:2005jk} induced by $\nu$-$\nu$ self-interactions in the deepest stellar
regions~\cite{Duan:2005cp,Duan:2006an,Hannestad:2006nj} 
(see, e.g.,~\cite{Duan:2010bg} for a recent review). 
The most important observational consequence of these effects is the ``swap'' of the $\nu_e$ and $\bar\nu_e$ spectra with
the non-electron $\nu_x$ and $\bar\nu_x$ spectra in certain energy
ranges~\cite{Duan:2006an,Hannestad:2006nj,Duan:2010bg,Fogli:2007bk,Fogli:2008pt, Raffelt:2007cb, 
Raffelt:2007xt, Duan:2007bt, Duan:2008za, Gava:2008rp, Gava:2009pj,Dasgupta:2009mg,%
Friedland:2010sc,Dasgupta:2010cd,Duan:2010bf,Mirizzi:2010uz}, 
 marked by the ``splits'',
which are the boundary features at the edges of each
swap interval.

One major challenge in the characterization of the self-induced effects on the SN neutrino 
flavor evolution  is related to
the the current-current nature of low-energy weak interactions, which implies 
a ``multi-angle term''~\cite{Duan:2006an}, $(1-{\bf v}_{\bf p} \cdot {\bf v}_{\bf q})$, where ${\bf v}_{\bf p}$
is the neutrino velocity~\cite{Qian:1994wh,Pantaleone:1992xh}.
Till recently,  all   studies  have assumed the 
axial symmetry for the  multi-angle kernel, reducing it to a ``multi-zenith-angle'' (MZA) term.
 In a non-isotropic medium this MZA term would not average to zero, producing a different 
refractive index for neutrinos propagating on different trajectories, something challenging the
maintenance of the collective effects~\cite{Raffelt:2007yz,EstebanPretel:2007ec,Sawyer:2008zs}.

Recently,  by means of a stability analysis of the linearized neutrino equations of motion~\cite{Raffelt:2013rqa}, 
 it has been pointed out that removing the  assumption of axial symmetry in the $\nu$ propagation, a new multi-azimuthal-angle (MAA) instability
 would emerge in the flavor evolution of the dense SN neutrino  gas in addition to the one caused by the MZA effects.
 Subsequently, the role of this instability   has been clarified in~\cite{Raffelt:2013isa,Duan:2013kba}, where it has been shown  with  simple toy models
that,  introducing small deviations in the assumed symmetries of the initial conditions
in a dense $\nu$ gas, new instabilities would be triggered.
The presence of MAA effects would unavoidably imply the breaking of
 the spherical symmetry in the flavor evolution after the onset of the conversions. 
 This would lead to a challenging multi-dimensional problem involving partial differential equations.
However, assuming that the variations of the global solution in the direction transverse to the radial ones
 are small, one can study the local solution  along a given line of sight, without 
worrying about the global behavior of the solution. 
This approach, even if  not completely self-consistent, allowed to obtain
the  first numerical solutions  of  the non-linear neutrino propagation equations in SN, 
introducing
the azimuthal angle as angular variable in addition to the usual zenith angle in the multi-angle kernel.
In the simulations   performed in~\cite{Mirizzi:2013rla},   
we considered simple energy spectra with an excess of  $\nu_e$ over $\bar\nu_e$.  
In these cases, we found  that even  starting with a complete axial symmetric neutrino emission, the MAA effects
would lead to significant flavor conversions in normal mass hierarchy (NH, $\Delta m^2_{\rm atm}>0$), in cases otherwise stable under the only 
MZA effects. Depending on the flavor asymmetry between $\nu_e$ and $\bar\nu_e$, MAA effects could lead to 
spectral swaps and splits (for large asymmetry) or to flavor decoherence (for small asymmetry). 

The purpose of this follow-up work is to take a closer
look to the MAA effects, considering realistic SN neutrino fluxes. 
We will perform at first a stability analysis of the linearized equations of motion, 
as described in~\cite{Raffelt:2013rqa,Banerjee:2011fj} and then we will compare these results with the numerical 
solution of the equations of motion. 
At this regard, we stress that  our numerical approach to the equations of motion is not
completely self-consistent,  since it is based on the assumption that the evolution can still
be characterized along a single radial line of sight. 
Therefore, our results should not be intended as  predictions for observable 
spectral signatures associated with the MAA effects. Rather, we are mostly interested in understanding
how the emergence of the MAA instability is associated with  the SN neutrino flux ordering.  
Since in the following we are interested in characterizing 
 the MAA instability,   we neglect the ordinary matter term, that could in principle suppress  the 
MZA~\cite{Chakraborty:2011nf,Chakraborty:2011gd,Saviano:2012yh,Sarikas:2011am}
 and the MAA~\cite{Chakraborty:2014nma} self-induced flavor conversions  at early times
 (at $t \lesssim 1$~s
after the core bounce).
We remind the reader that SN $\nu$ energy
spectra present crossing points  in the energy variable (energies
where $F_{\nu_e} = F_{\nu_x}$, and  $F_{\bar\nu_e} = F_{\bar\nu_x}$) and 
can develop instabilities around them.
It is known that the impact of the MZA effects on the flavor evolution crucially depends 
on the crossing pattern~\cite{Dasgupta:2009mg,%
Friedland:2010sc,Dasgupta:2010cd,Duan:2010bf,Mirizzi:2010uz}. At this regard we find important to evaluate how the MAA effects
impact the SN $\nu$ fluxes for different crossing configurations. 

In this context, during early post-bounce times (especially during the accretion phase, at post-bounce times
$t_{\rm pb} \lesssim 0.5$~s)
one expects    fluxes with a strong flavor hierarchy 
$N_{\nu_e} \gg N_{\bar\nu_e} \gg N_{\nu_x}$, defined in terms
of the total neutrino number fluxes $N_\nu$ for the different
flavors. These would practically correspond to
neutrino spectra with a single crossing point at $E \to \infty$ 
 since $F_{\nu_e}(E) >
F_{\nu_x}(E)$ for all the relevant energies (and analogously
for $\bar\nu$).
For spectra with a single crossing and with an excess of $\nu_e$ over $\bar\nu_e$ we find that
the MAA effect would lead to a new flavor instability in NH. With our numerical approach we find
  a splitting configuration
in the oscillated $\nu$ fluxes   rather close 
to what produced by the bimodal instability in inverted mass
hierarchy (IH, $\Delta m^2_{\rm atm}<0$) in the presence of only MZA effects. 

At later times, during the cooling phase (at $t_{\rm pb} \gtrsim 1$~s),  one instead expects a moderate flavor hierarchy
$N_{\nu_e} \gtrsim N_{\bar\nu_e} \gtrsim N_{\nu_x}$,
where a ``cross-over'' among nonelectron
and electron species is possible. This
case would correspond to neutrino spectra with multiple
crossing points, i.e. with $F_{\nu_e}(E) > F_{\nu_x}(E)$ at lower
energies, and $F_{\nu_e}(E) < F_{\nu_x}(E)$ at higher energies (and
analogously for $\bar\nu$).
It has been shown in the axial symmetric case that  the presence of MZA effects would inhibit flavor
conversions  otherwise possible at low-radii in a single-zenith-angle (SZA) scheme~\cite{Duan:2010bf,Mirizzi:2010uz}.
MAA effects that  in  SZA scheme would lead to a  flavor instability, 
are also inhibited by the MZA suppression. Therefore, when flavor conversions start at large radii they 
do not have enough time to grow. As a result multiple spectral splits  appear in both normal and inverted hierarchy
as found in the axial symmetric case. 

These results are presented as the following. 
In Sec.~2 we introduce the  neutrino equations of motion without 
imposing axial asymmetry in the neutrino-neutrino interaction term. 
Then, we describe  the setup to perform the  stability 
analysis of the linearized equations of motion. 
In Sec.~3 we present
different models for supernova neutrino fluxes with different crossing configurations.
For these models we perform the stability analysis and we compare the results
with   the  numerical flavor evolution. Finally in  Sec.~4 we summarize our results and we conclude.

\section{Setup of the flavor evolution}

\subsection{Equations of motion}
 
 The characterization of the global solution for neutrino flavor evolution 
 once the axial symmetry is broken by the MAA effects,   would
in general  require 
to  take into account also variations along the  transverse
direction to the neutrino propagation one. This would make  
the solution of the equations of motion extremely challenging .
However, 
\emph{before} the spherical symmetry is broken, one can still characterize the flavor evolution
 along a given line of sight.
In order to make the problem  treatable also after the axial symmetry breaking,  in our study we will use as
 \emph{working hypothesis} that transverse variations
remain small for all the flavor evolution. Given this limitation, the results
obtained with our simulations should not  be intended to provide  a benchmark to characterize detailed  spectral features.
Rather, they would justify more advanced analysis to understand if the features found here are present also 
in more realistic descriptions. 
In our study we will focus  mostly on  the question of the emergence of an   instability
in the dense neutrino gas under MAA effects,
by finding the possible onset for the flavor conversions. 
This latter will be at first determined with the stability analysis and then compared with the approximate
numerical solution.

Under our simplified assumption,  the flavor evolution  depends only on $r$, $E$ and ${\bf v}_{\bf p}$.
 Following~\cite{Banerjee:2011fj, Raffelt:2013rqa}, we write the equations of motion for the flux matrices  $\Phi_{E,u, \varphi}$  as function of the radial coordinate.
 The diagonal $\Phi_{E,u, \varphi}$ elements represent
the ordinary number fluxes 
integrated
over a sphere of radius $r$.
We normalize the flux matrices to the  ${\overline\nu}$ number flux at the neutrino-sphere
 radius $R$.
 The off-diagonal elements,
which are initially zero, carry a phase information associated with
flavor mixing.
 We use negative energy $E$  for anti-neutrinos.
Concerning the angular variables, we parameterize every zenith angular mode in terms of the variable 
$u=\sin^2\theta_R$, where
  $\theta_R$ is the emission angle relative to the radial direction
of the neutrinosphere~\cite{EstebanPretel:2007ec,Banerjee:2011fj}, 
while $\varphi$ is the azimuth angle of the neutrino velocity ${\bf v_p}$. 
For simplicity, we neglect possible residual scatterings that could affect $\nu$'s after the 
neutrinosphere, producing a small ``neutrino halo'' that would broaden the $\nu$ angular 
distributions~\cite{Cherry:2012zw,Sarikas:2012vb}.
Then, the equations of motion read~\cite{Banerjee:2011fj,Sigl:1992fn}
\begin{equation}
\textrm{i}\partial_r \Phi_{E,u, \varphi}=[H_{E,u, \varphi},\Phi_{E,u,\varphi}] \,\ ,
\label{eq:eom1}
\end{equation}
with the Hamiltonian~\cite{Qian:1994wh,Banerjee:2011fj,Sigl:1992fn}
 \begin{eqnarray}
& & H_{E,u} = \frac{1}{v_{u}}\frac{M^2}{2E}  \nonumber \\
 &+& \frac{\sqrt{2}G_F}{4\pi r^2} \int d \Gamma_{E,u, \varphi}^{\prime}  \left(\frac{1-v_{u}v_{u^\prime}-{\bm\beta}\cdot {\bm\beta}^{\prime}}
{v_{u}v_{u^\prime}} \right)  \Phi^\prime \,\ .
 \label{eq:eom2}
 \end{eqnarray}
The matrix $M^2$ of neutrino mass-squares causes vacuum
flavor oscillations. We work in a two-flavor scenario,
where we take as  mass-square difference the atmospheric one,  $\Delta m^2_{\rm atm}= 2 \times 10^{-3}$~eV$^2$.
When needed, we assume  a small (matter suppressed) in-medium mixing $\Theta = 10^{-3}$. 
The term at the second line represents
the $\nu$-$\nu$ refractive effect, where 
 $\int d \Gamma_{E,u, \varphi} = \int_{-\infty}^{+\infty}d E
 \int_{0}^{1}du  \int_{0}^{2\pi}d\varphi$.  In the multi-angle kernel, 
 the radial velocity of a mode with angular label $u$ is
$v_{u} = (1-u R^2/r^2)^{1/2}$~\cite{EstebanPretel:2007ec} and the transverse velocity is 
$\beta_{u}= u^{1/2} R/r$~\cite{Raffelt:2013rqa}.
The term 
$ {\bm\beta}\cdot{\bm\beta}^{\prime} = \sqrt{u u^{\prime}}{R^2/r^2}
\cos(\varphi-\varphi^{\prime})$ is the 
responsible for the breaking of the axial symmetry.

As in~\cite{Mirizzi:2013rla}, to  solve numerically  Eq.~(\ref{eq:eom1})  we use an integration routine for stiff ordinary differential equations taken
from the NAG libraries~\cite{nag} and based on an adaptive method.
We have used  $N_\varphi=40$ modes for $\varphi \in [0;2\pi]$, $N_u=100$
for $u \in [0;1]$ and  $N_\omega=80$ for $E \in [0;80]$~MeV.

\subsection{Stability conditions}

In order to perform the stability analysis, we  linearize the equations of motion
[Eq.~(\ref{eq:eom1})-(\ref{eq:eom2})],  following the approach of~\cite{Banerjee:2011fj,Raffelt:2013rqa} 
to which we address the interested reader for further details.
 We   write the flux matrices in the form
\begin{equation}
\Phi_{\omega,u}= \frac{\textrm{Tr}\Phi_{\omega,u, \varphi}}{2}+
\frac{g_{\omega,u, \varphi}}{2}
\left( \begin{array}{cc} s &  S \\
S^{\ast} & -s
\end{array} \right) \,\ .
\label{eq:pHI}
\end{equation}
Here, instead of energy,  we use   the 
frequency variable $\omega= \Delta m^2_{\rm atm}/2E$, and  we introduce the 
neutrino flux difference distribution $g_{\omega,u, \varphi}\equiv g(\omega,u, \varphi)$.
This is the dimensionless spectrum 
representing ${F^R_e-F^R_x}$ at the neutrinosphere,
normalized  to the ${\overline\nu}$ flux. 
In the following we will always assume axial symmetry of the neutrino emission. Therefore $ g(\omega,u, \varphi)=  g(\omega,u)/2\pi$.
Self-induced flavor transitions  start when the off-diagonal term $S$ grows
 exponentially.
 One can write the solution of the linearized  evolution equation (see~\cite{Banerjee:2011fj,Raffelt:2013rqa})
in the form $S = Q _{\omega,u, \varphi} e^{-i\Omega r}$ with complex frequency
$\Omega= \gamma + i \kappa$ and eigenvector $Q _{\omega,u, \varphi}$. A solution with 
$\kappa >0$ would indicate an exponential increasing $S$, i.e. an instability.
Then the solution   can  be recast in the form of an eigenvalue equation
for $Q _{\omega,u, \varphi}$, from which one gets a set of  consistency conditions to be satisfied.
These conditions can be expressed in terms of the following quantities
\begin{eqnarray}
J_n &=& \int d\omega \,du\, g_{\omega,u} u^n \frac{\omega -\omega^*}
{(\omega -\omega^*)^2 + \kappa^2} \,\ , \nonumber \\
K_n &=& \int d\omega \, du\, g_{\omega,u} u^n \frac{\kappa}
{(\omega -\omega^*)^2 + \kappa^2} \,\ ,
\label{eq:consit}
\end{eqnarray}
where we introduced the \emph{resonant frequency}
\begin{equation}
\omega^*(\gamma)= \gamma - u\epsilon \mu  \,\ ,
\label{eq:res}
\end{equation}
in terms of the $\nu-\bar\nu$ asymmetry parameter 
$\varepsilon = \int d\Gamma g$, where
$\int d \Gamma = \int_{-\infty}^{+\infty}d \omega
 \int_{0}^{1}du  \int_{0}^{2\pi}d\varphi$, and of
 the strength of the $\nu$-$\nu$ interaction
\begin{eqnarray}
\mu &=& \frac{\sqrt{2}G_F (N_{{\bar\nu}_e}-N_{{\bar\nu}_x})}{4 \pi r^2}\frac{R^2}{2 r^2} \nonumber \\
&=& \frac{3.5 \times 10^{5}}{r^4} 
\left(\frac{L_{\bar\nu_e}}{\langle E_{\bar\nu_e}\rangle} -
\frac{L_{\bar\nu_x}}{\langle E_{\bar\nu_x}\rangle}
 \right)
\frac{15 \,\ \textrm{MeV}}{10^{51} \,\ \textrm{erg}/\textrm{s}}
 \left(\frac{R}{10 \,\ \textrm{km}} \right)^2 . \nonumber
\end{eqnarray}
In the case of axial-symmetric case,  
 one arrives
at two real equations~\cite{Banerjee:2011fj}
\begin{eqnarray}
(J_1-\mu^{-1})^2 &=& K_1^2 + J_0 J_2 -K_0 K_2 \,\ , \nonumber \\
(J_1-\mu^{-1}) &=& \frac{J_0 K_2 + K_0 J_2}{2 K_1} \,\ .
\label{eq:JnKn}
\end{eqnarray}
When these equations admit a solution, they determine the well-known \emph{bimodal} instability.
In the non-axial-symmetric another consistency condition arises, that leads to 
these two additional equations 
\begin{eqnarray}
J_1+\mu^{-1} &=& 0 \,\ , \nonumber \\
K_1 &=& 0 \,\ ,
\label{eq:JnKnMAA}
\end{eqnarray}
that characterize the new \emph{multi-azimuthal-angle} instability. 

A flavor instability is present whenever Eqs.~(\ref{eq:JnKn})--(\ref{eq:JnKnMAA}) admit a  solution $(\gamma, \kappa)$. 
When an instability occurs, for a given  angular mode $u_0$ the function 
$|Q _{\omega,u_0}|$ is a Lorentzian~\cite{Dasgupta:2009mg}, centered around $\omega^*$,  with a width $\kappa$. 
Finally, we remind that the previous consistency equations are for the inverted mass hierarchy (IH). 
In  normal mass hierarchy (NH) one should simply change the sign of $\omega \to -\omega$
in Eqs.~(\ref{eq:JnKn})--(\ref{eq:JnKnMAA}).

\section{Multi-azimuthal-angle effects for different supernova models}

\begin{figure}[!t]
\begin{center}
 \includegraphics[angle=0,width=0.4\textwidth]{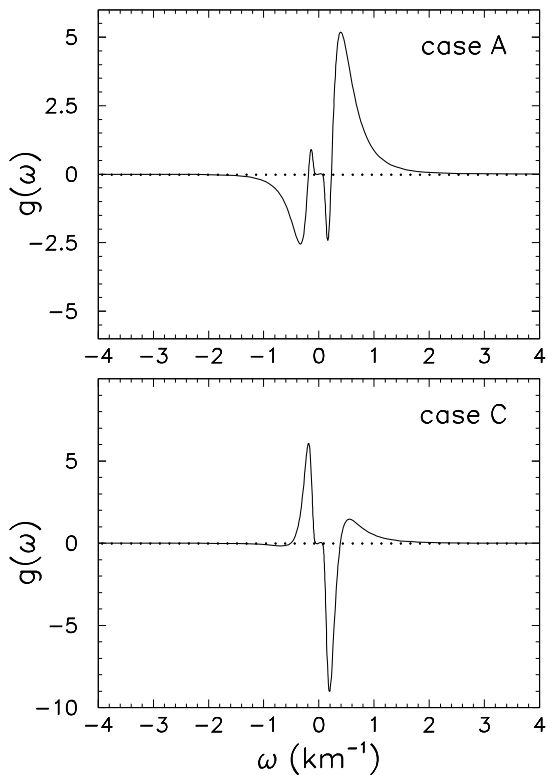} 
    \end{center}
\caption{Spectrum $g(\omega)$ for the case  $\cal A$ (upper panel) and $\cal C$
(lower panel).} \label{fig1}
\end{figure}

\subsection{Supernova flux models}

In order to study the MAA effects of SN neutrino flavor conversions, we fix our benchmark models
for the initial SN neutrino fluxes at the neutrinosphere. 
We  factorize the $\nu$ flux of each flavor as  
$F_{\nu_\alpha}(E)= N_{\nu_\alpha} \times \varphi_{\nu_\alpha}(E).$
The $\nu$ number $N_{\nu_\alpha}=L_{\nu_\alpha}/{\langle E_{\nu_\alpha}\rangle}$ is expressed in terms of
the $\nu$ luminosity  $L_{\nu_\alpha}$ and of  the $\nu$ average energy ${\langle E_{\nu_\alpha}\rangle}$ of the different species.
The function $\varphi_{\nu_\alpha} (E)$ is the normalized $\nu$ energy spectrum ($\int dE \varphi_{\nu_\alpha} (E)=1$). 
We parametrize the energy spectrum  as in~Ref.~\cite{Raffelt:2003en,Tamborra:2012ac}
\begin{equation}
\varphi(E)=\frac{(1+\alpha)^{1+\alpha}}{\Gamma(1+\alpha)}\frac{E^\alpha}{\langle E_\nu\rangle^{\alpha+1}}\exp\left[-\frac{(1+\alpha)\,E}{\langle
E_\nu\rangle}\right]\, ,
 \label{eq:varphi}
\end{equation}
where we fix the spectral parameter to  $\alpha=3$ for all species.
We fix the neutrino average energies at
\begin{equation}
(\langle E_{\nu_e}\rangle, \langle E_{{\bar\nu}_e}\rangle, \langle E_{\nu_x}\rangle)=(12, 15, 18)\,\  
\textrm{MeV} \,\ .
\end{equation}
Concerning the possible $\nu$ flux ordering we consider two cases, widely used as benchmark models in the previous 
literature~\cite{Dasgupta:2009mg,%
Friedland:2010sc,Dasgupta:2010cd,Duan:2010bf,Mirizzi:2010uz}.  As representative of the accretion phase (labelled as case ${\cal A}$), we take
\begin{equation}
N_{\nu_e}:N_{\bar\nu_e}:N_{\nu_x}=1.50:1.00:0.62 \,\ ,
\end{equation}
corresponding to an asymmetry paramter $\epsilon = 1.34$.
In this case, we choose as neutrino luminosities (in units of $10^{51}$ erg/s) 
\begin{equation}
L_{\nu_e} = 2.40 \,\ , \,\  L_{\bar \nu_e} = 2.00 \,\ , \,\ L_{\nu_x} = 1.50 \,\ . 
\end{equation}

Instead, as representative of the cooling phase, dubbed  case ${\cal C}$,
we choose
\begin{equation}
N_{\nu_e}:N_{\bar\nu_e}:N_{\nu_x}=1.13:1.00:1.33  \,\ ,
\end{equation}
giving $\epsilon = 0.37$. In this case we take as neutrino luminosities (in units of $10^{51}$ erg/s) 
\begin{equation}
L_{\nu_e} = 1.20 \,\ , \,\  L_{\bar \nu_e} = 1.34 \,\ , \,\ L_{\nu_x} = 2.14 \,\ . 
\end{equation}

In Figure~1  we show the function $g_{\omega}= \int du\,  g_{\omega,u}$ for the case
${\cal A}$ (upper panel) and for the case ${\cal C}$ (lower panel). Note that here and in the following
we will assume   half-isotropic $\nu$ angular distributions
(i.e. with  all outward-moving angular modes 
equally occupied and all the backward-moving modes
empty). 
Both spectra present three crossing points, where  
 $g_{\omega}=0$, so naively one would expect similar instability conditions in these two cases. However, as we will see
in the following, the instabilities in these two cases are significantly different.

\subsection{Case ${\cal A}$}

The $g_{\omega}$ function for the case ${\cal A}$ (Fig.~\ref{fig1} upper panel),
 presents three crossing points in the $\omega$ variable.
According to what explained in~\cite{Dasgupta:2009mg} and confirmed by the stability analysis in~\cite{Banerjee:2011fj}
in the axial symmetric case, the bimodal  instability is  associated to crossings
with positive slope in IH  and to 
crossings with negative slope  in NH. 
 However, as discussed in~\cite{Dasgupta:2009mg} and observed in many numerical simulations,
 a narrowly spaced triple crossing
can superficially act like a single one at $\omega=0$. Therefore, 
the bimodal instability would occur  in IH and produce no effect in
NH. 
If one breaks the axial symmetry, allowing for MAA effects, in the stability analysis of~\cite{Raffelt:2013rqa}
and from the numerical simulations of~\cite{Mirizzi:2013rla} one finds that the positive crossings would be unstable in NH 
and the negative ones in IH. Therefore, due to the narrowness of the three crossings in the   $g_{\omega}$,
we expect that MAA effect would produce an  instability in NH and no new effect in IH. 
In Fig.~2 we show the radial evolution of the $\kappa$ function for $\nu$ fluxes of the case 
 ${\cal A}$, obtained solving Eqs.~(\ref{eq:JnKn})--(\ref{eq:JnKnMAA}).
 Left panel refers to the NH, while right panel to the IH one. 
We consider both MZA case (continuous curves) with half-isotropic angular distributions, and
single-zenith-angle (SZA) case (dashed curves), where we took $g(u)=\delta(u-u_0)$ with $u_0=0.5$. 
We realize that both the configurations are unstable.
As explained before, the instability in NH is associated 
with MAA effects [Eqs.~(\ref{eq:JnKnMAA})]. We find that in the SZA case
the peak of  $\kappa \simeq 1.8$ is reached at $r \simeq 85$~km. In the MZA case this is 
slightly suppressed at $\kappa \simeq 1.4$.
The  curves of $\kappa$ in IH, associated with the bimodal instability [Eq.~(\ref{eq:JnKn})]
are rather similar to what discussed for the NH.
 At this regard, the symmetry between  the MAA instability   in NH and
the bimodal one in IH has been discussed in~\cite{Raffelt:2013isa}.

\begin{figure*}[!t]
 \includegraphics[angle=0,width=0.8\textwidth]{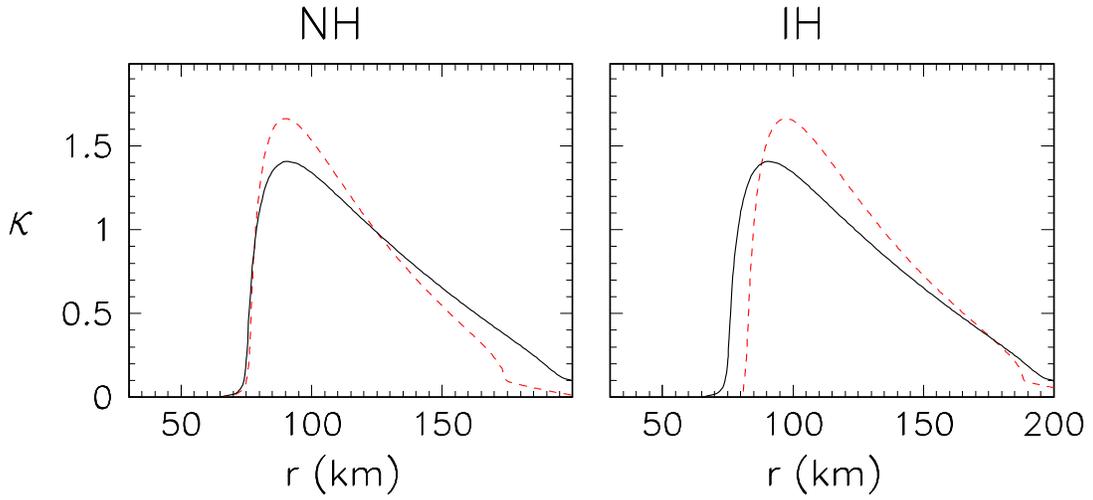} 
\caption{Case ${\cal A}$. Radial evolution of the $\kappa$ function in NH (left panel) and 
IH (right panel). Continuous curves correspond to the MZA case while dashed ones to the SZA case.
The growth of  $\kappa$ in the NH is associated to the MAA instability, while in IH to the bimodal
 instability. 
} \label{fig2}
\end{figure*}

\begin{figure*}[!t]
 \includegraphics[angle=0,width=0.8\textwidth]{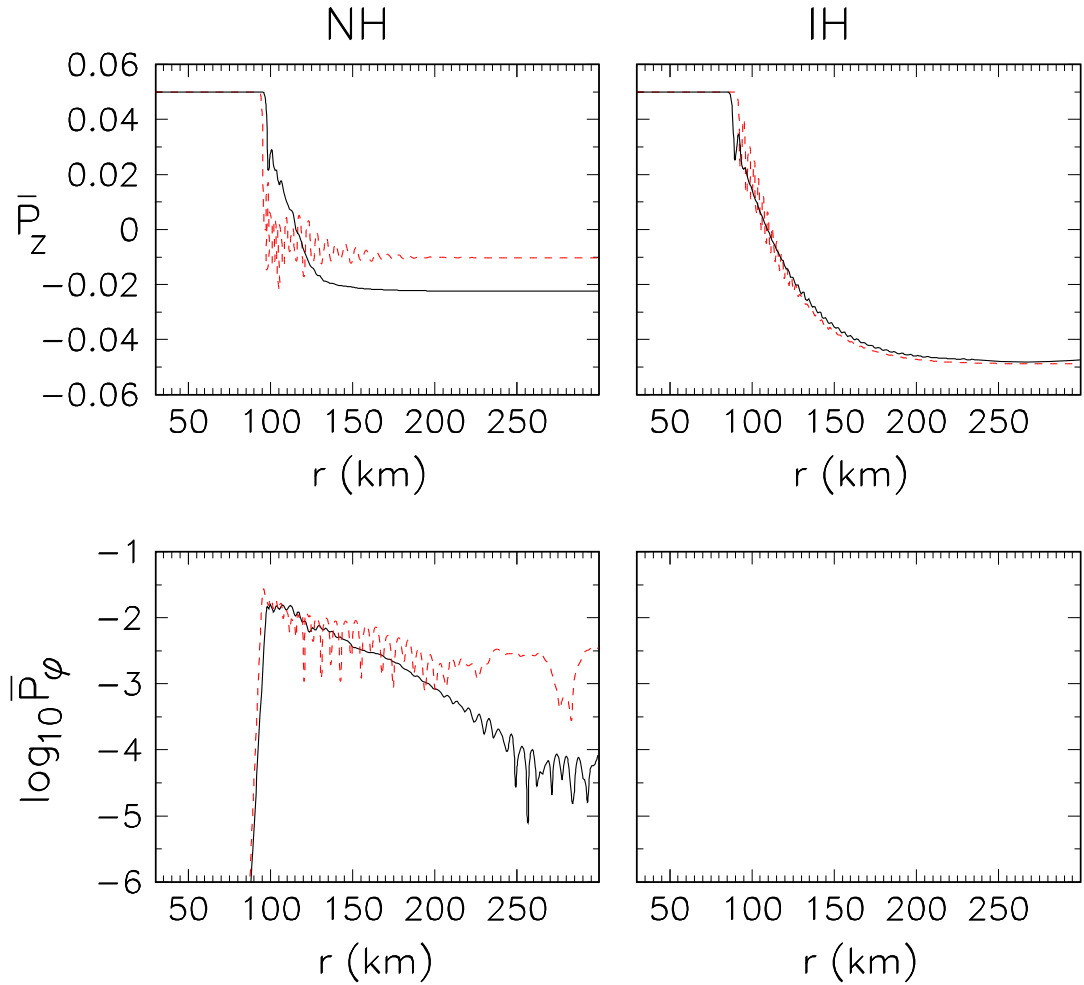} 
\caption{Case ${\cal A}$. Upper panels: Radial evolution of the integrated z-component 
${\bar P}_z$ of the polarization vector  for $\bar\nu$. 
Lower panels: Radial evolution of the integrated dipole
${\bar P}_\varphi$  for $\bar\nu$. 
Left panels refer to normal hierarchy, right panels to inverted hierarchy.
Dashed curves are for the MAA, SZA evolution; while
continuous curves are for the MAA, MZA case.
} \label{fig3}
\end{figure*}

In the upper panels of Figure~3 we represent the radial evolution of the  $z$-component
${\bar P}_z$ of the $\bar\nu$ polarization vector  ${\bf P}_{\omega, u, \varphi}$
integrated over $\omega$, $\varphi$ and $u$ for neutrino fluxes in the ${\cal A}$ case. 
As known, this quantity is related to the flavor content of the $\nu$ ensemble. 
Left panel refer to NH and right panel to IH. 
We compare  single-zenith-angle (SZA) evolution (dashed curves) with $u_0=0.5$,  with the complete MZA and MAA cases
(continuous curves).
In IH we observe the usual flavor evolution, found  in the axial symmetric case. 
 In particular, we have 
 synchronized oscillations~\cite{Pastor:2001iu}   till $r \simeq 85$~km, where
 all the polarization vectors of the different modes stay pinned to their original values.
 The onset of the flavor conversions is in agreement with what found with the stability analysis.
Significant flavor conversions start only after synchronization
when pendular oscillations~\cite{Hannestad:2006nj}, triggered by the bimodal instability~\cite{Samuel:1996ri},  start to swap the
flavor content of the system. In particular, ${\bar P}_z$ is completely swapped at the end of the 
flavor evolution. 
MAA and
MZA effects do not play a significant role, since one observes a  ``quasi-single angle''
behavior~\cite{EstebanPretel:2007ec}, driven by the only bimodal instability.
The NH case, that would have been stable in the axial symmetric case, now presents
relevant flavor conversions  at $r \gtrsim 95$~km, analogous to what   recently found for the simple spectra considered in~\cite{Mirizzi:2013rla}. Again we find an agreement between the onset of the flavor conversions found numerically
and what predicted with the stability analysis. 
In this case, the inversion of ${\bar P}_z$  is not complete and there is also a slight difference
between the SZA and MZA case. 

In order to track the growth of the MAA instability, in the lower panels of Figure~3 we plot
the dipole term in $\cos \varphi$ of the transverse components of the antineutrino integrated polarization vector~\cite{Mirizzi:2013rla}
\begin{equation}
{\bar P}_\varphi = \left\{\frac{1}{2\pi} \int d\Gamma \,\   g(\omega) [({\bar P}^{x}_{\omega,u, \varphi})^2
+ ({\bar P}^{y}_{\omega,u, \varphi})^2] \cos \varphi \right\}^{1/2}\,\ .
\end{equation}
From the Figure one realize that while in IH case, this function is always null, confirming that no
azimuthal dependence in the polarization vectors in generated, in NH ${\bar P}_\varphi$ exponentially grows
reaching the peak at  $r \gtrsim 95$~km where flavor conversions are triggered by the MAA effects. 
The rise  of  ${\bar P}_\varphi$ in SZA (dashed curve) and MZA (continuous curve) case is rather similar.

\begin{figure*}[!t]
 \includegraphics[angle=0,width=0.8\textwidth]{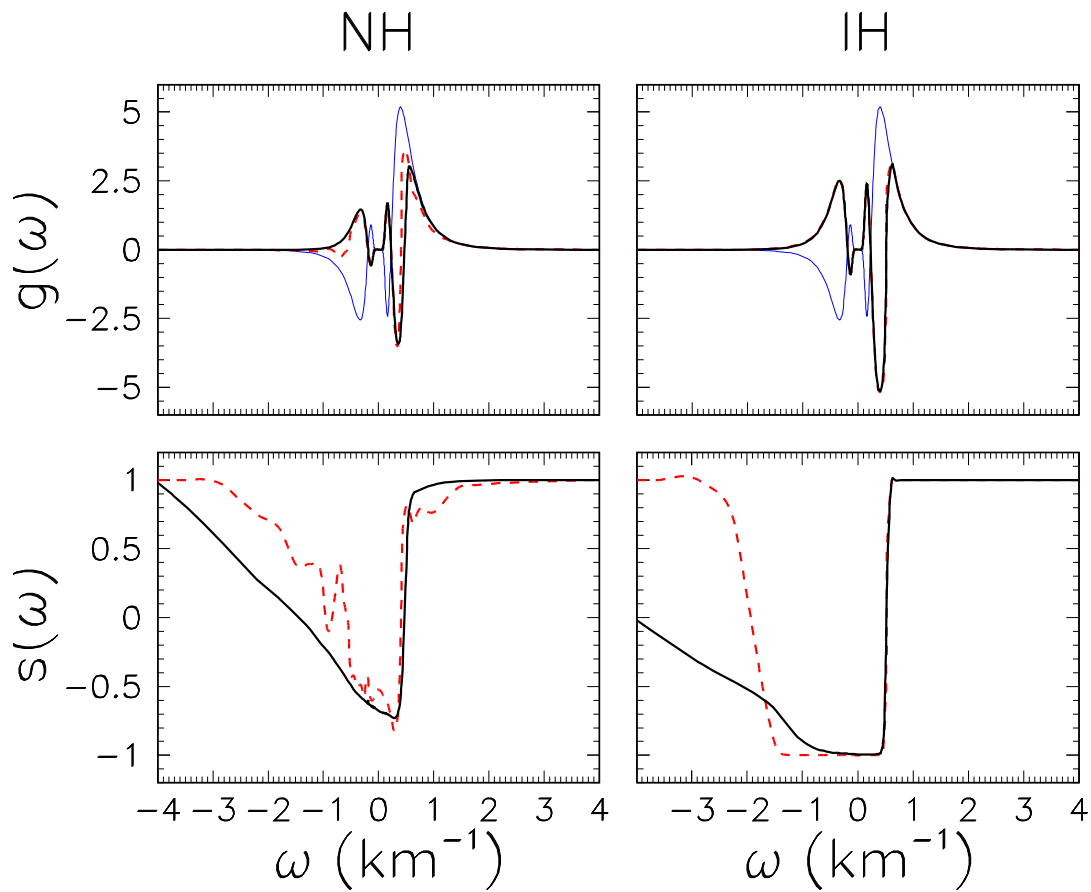} 
\caption{Case ${\cal A}$.  Upper panels: initial  (thin continuous curves) and final $g(\omega)$
for  the MAA evolution
in SZA (dashed curves) and MZA (thick continuous curves) case
 for normal hierarchy (left panel) and inverted hierarchy (right panel).
Lower panels: swap function, i.e. ratio of final with initial spectra 
in SZA  (dashed curves) and MZA (thick continuous curves).
} \label{fig4}
\end{figure*}

\begin{figure*}[!t]
 \includegraphics[angle=0,width=0.8\textwidth]{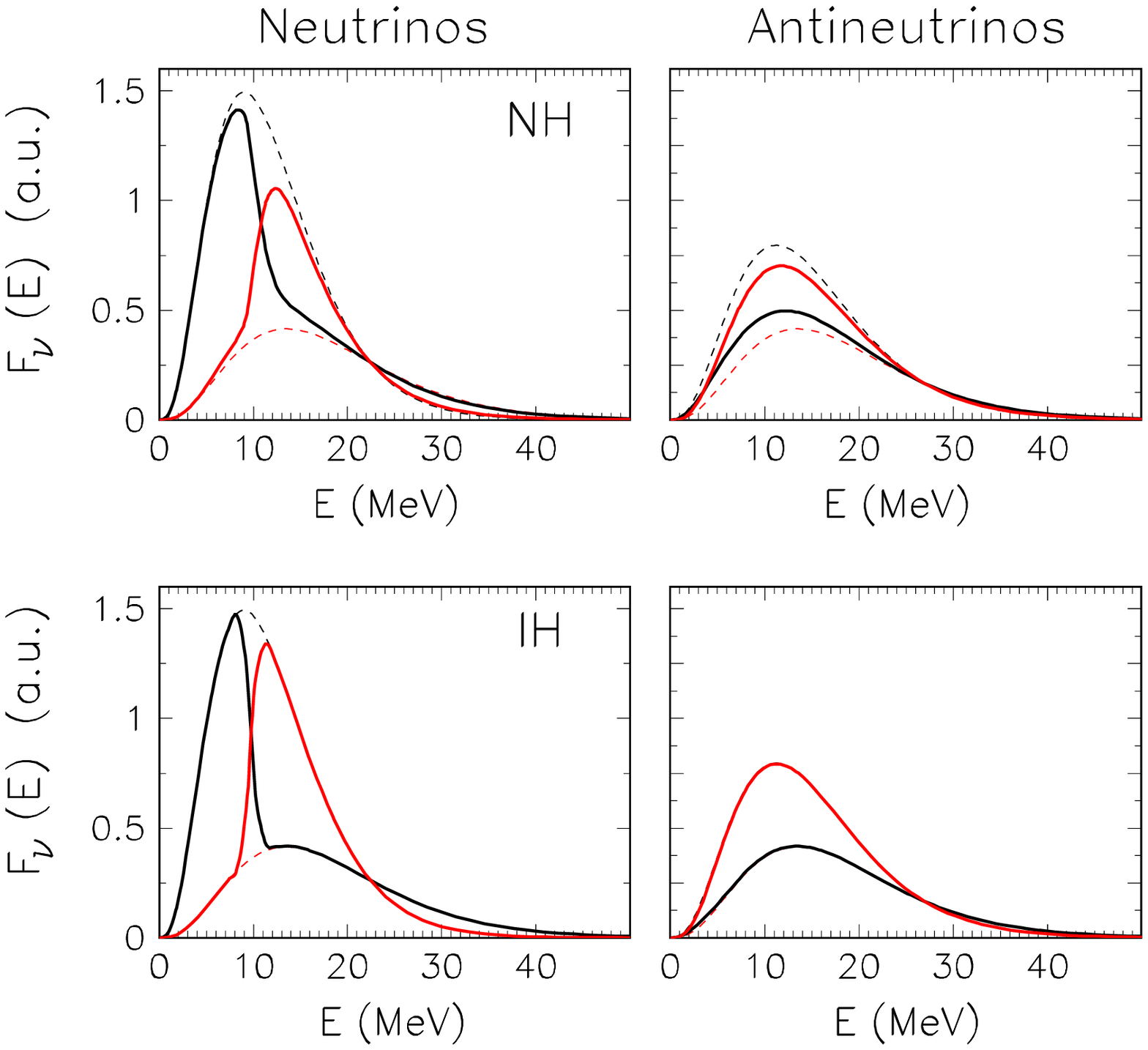} 
\caption{Case ${\cal A}$. MZA and MAA flavor evolution for $\nu$'s
(left panel) and $\bar\nu$'s (right panel) in NH (upper panels) and IH (lower panels).
 Energy spectra initially for
$\nu_e$ (black dashed curves) and $\nu_x$ (light dashed curves)
and after collective oscillations for $\nu_e$ (black continuous curves)
and $\nu_x$ (light continuous curves).
} \label{fig5}
\end{figure*}

For completeness, we find interesting to show
how the MAA induced flavor conversions 
 affect the SN neutrino fluxes in our simplified numerical scheme. 
As already noticed, the $g_{\omega}$ function for the case ${\cal A}$ looks as
a spectrum with a single crossing at $\omega=0$.
Therefore, the bimodal instability would produce 
 a broad swap around $\omega = 0$ in IH, while the MAA instability would produce the swap in NH. 
Our results are shown in Figure~4. In the upper panels we show the initial $g(\omega)$ (light continuous curves),
the final one for the MAA flavor evolution in the SZA case (dashed curve) and the final one in the MZA case (thick continuous curves).
In the lower panels we represent the swap function $s(\omega)$ given by the ratio of the final with
respect to the initial $g(\omega)$
for the SZA case (dashed curves) and  the MZA case (thick continuous curves).
Left panels refer to the NH case, while right panels to IH. 
We remark that the final spectra in NH and IH are rather  close. 
 As expected, the MAA instability in NH produces  a split at $\omega \simeq 0.4$~km$^{-1}$, while  in IH the split induced  by
the bimodal instability is  at $\omega \simeq 0.5$~km$^{-1}$.  
 In the SZA case another split would appear in the $\bar\nu$ at $\omega \simeq -0.5$~km$^{-1}$ in NH,
 and at $\omega \simeq -1.5$~km$^{-1}$ in IH.
 As noticed in~\cite{Mirizzi:2013rla}, both these $\bar\nu$ splits are smeared by MZA effects.
 The presence of spectral splits and swaps lead by the MAA instability in the case ${\cal A}$ is consistent with
 what presented  in~\cite{Mirizzi:2013rla}, where an ordered behavior in the final $\nu$ spectra under MAA effects was found 
 for large flavor asymmetries ($\varepsilon \gtrsim 1$). 
 Moreover, the symmetry between the effects of the  bimodal instability in IH and the MAA one in NH has been
recently pointed out with a simple toy-model in~\cite{Raffelt:2013isa}.

Finally, in Figure~5 we show the intial (dashed curves) and final fluxes (continuous curves) for 
$e$ (black curves) and $x$ (light curves)  flavors, for neutrinos (left panels) and antineutrinos (right panels).
We consider both MZA and MAA effects.
Upper panels show the NH while lower panels refer to IH. 
We realize that the final spectra are remarkably similar in both the hierarchies. 
In particular, neutrino spectra present a split at $E\simeq 12$~MeV. The split is sharper in IH where
it is driven by the bimodal instability. 
The $\bar\nu$ spectra are swapped with respect to the initial ones. In particular, as expected from the 
swap function of Figure~3, the swap is more complete in IH, where deviations from a complete swap are confined
to low energies ($E\lesssim 2$~MeV).

\subsection{Case ${\cal C}$}

The $g_{\omega}$  spectrum for the case  ${\cal C}$ (Fig.~1 lower panel)  has been widely studied in the axial symmetric case
as  representative of energy spectra with a flux ordering
$N_{\nu_x} \gtrsim N_{\nu_e} \gtrsim N_{\bar\nu_e}$ and energy spectra 
 with multiple crossing points  leading to multiple 
splits around these points (see, e.g.,~\cite{Dasgupta:2009mg,Friedland:2010sc,Dasgupta:2010cd,Duan:2010bf,Mirizzi:2010uz}). 
Remarkably, with spectra of type  ${\cal C}$ one expects
both the hierarchies to be unstable under MAA and bimodal  effects, since spectra present
both  positive (unstable in NH for MAA effects and in IH for the bimodal ones) 
and  negative  (unstable in IH for MAA effects and in NH per the bimodal ones) crossings. 
In Figure~6 we show the radial evolution of the function $\kappa$, obtained solving Eqs.~(\ref{eq:JnKn})--(\ref{eq:JnKnMAA}).
Left panels refer to NH, while right ones to IH.
Differently from the case ${\cal A}$, these two equations will have simultaneous solutions in both the 
hierarchies. Therefore, in the upper panels we show the solutions of Eqs.~(\ref{eq:JnKn}) [bimodal instability], 
while in  lower panels  refer to  Eqs.~(\ref{eq:JnKnMAA}) [MAA instability].
As in Fig.~3, continuous curves refer to  MZA, while dashed ones are for  SZA. 
We remark that  Eqs.~(\ref{eq:JnKn}) have a couple of solution $(\gamma, \kappa)$ corresponding
to positive and negative $\gamma$ respectively (see, e.g.,~\cite{Banerjee:2011fj}).
However,  in the following we will show only  the case with the  larger $\kappa$. 
 Starting from the NH case, we see that in the SZA situation both bimodal and MAA effects
  produce a sharp rise in $\kappa$ around $r\simeq 40$~km. 
 Therefore, low-radii conversions would be possible in these cases. 
 Since both the instabilities are comparable (peak value of $\kappa \simeq 1.2$), one expects 
that both would equally contribute to the further flavor evolution. 
Passing now to the MZA case, as expected from~\cite{Banerjee:2011fj}, the instabilities are strongly suppressed with respect 
to the previous cases. Moreover, the peaks are shifted at larger $r$. 
In particular, for the bimodal instability the peak is at $r \simeq 80$~km
with a value $\kappa \simeq 0.7$, while  the MAA instability has the peak at 
$r \simeq 60$~km with a value  $\kappa \simeq 0.3$. The strong suppression of the MAA
instability seems to suggest that this one would play a sub-leading role in the flavor 
evolution in the MZA situation. Passing to the IH, we observe a similar trend. 
Namely, in SZA both the bimodal and MAA instabilities lead to comparable effects. 
In the SZA case, the bimodal  instability  has a peak 
$\kappa \simeq 0.7$ at $r \simeq 80$~km, while the MAA one is smeared and more suppressed
with a peak $\kappa \simeq 0.5$ at $r \simeq 90$~km.

In Figure~7 we represent the radial evolution of the integrated vectors ${\bar P}_z$ and
${\bar P}_{\varphi}$ in the same format of Fig.~2.
As expected from the stability analysis, we see that in this case the flavor evolution in SZA and MZA case is remarkably different. 
In particular, in the SZA case  we see that
differently from the case ${\cal A}$, flavor
conversions are possible at low radii ($r \gtrsim 40$~km) in a
region where we would have expected synchronization, in agreement with the results
of the stability analysis. 
This effect has been discussed in literature: 
in the presence of multiple crossing points it is possible 
the flip of  parts of the original neutrino spectra around
the crossing points~\cite{Raffelt:2008hr}. Therefore, the synchronization
behavior found for single-crossed spectra is
not necessarily stable in this case. Indeed, in~\cite{Raffelt:2008hr}
it has been discussed the possibility of a novel form
of flavor conversions for large $\mu$ in terms of a self-induced
parametric resonance that would destabilize the
synchronization. In this case
also the MAA instability can grow at relatively low-radius in both the mass hierarchies.
From the plot of ${\bar P}_{\varphi}$, one see that the peak is reached 
at $r \gtrsim 60$~km in both NH and IH. 
 
Passing to the MZA case  of Fig.~7, we find that
now the low-radii flavor conversions are suppressed, as predicted from the stability analysis. 
This
effect has been  described in~\cite{Raffelt:2008hr}, where  it
was  pointed the connection between this behavior and
the MZA  suppression of the parametric resonance.
Indeed, as understood from previous literature~\cite{Duan:2010bf,Mirizzi:2010uz}, MZA effects
introduce a large dispersion in the neutrino-neutrino
potential, that prevents any possible collective flavor
conversion at low radii. 
The MZA suppression implies that flavor conversions in this case 
do start only at $r \gtrsim 80$~km. 
Remarkably, also MAA effects are suppressed at low-radii and when flavor conversions
start they do not have enough time to  grow. 
Indeed, we have seen with the stability analysis 
that the MAA instability is more suppressed with respect to the  bimodal
one. 
Indeed ${\bar P}_{\varphi}$ remains identically
null. In conclusion in this case, the flavor evolution is governed by the only bimodal and
MZA instabilities.

\begin{figure*}[!t]
 \includegraphics[angle=0,width=0.8\textwidth]{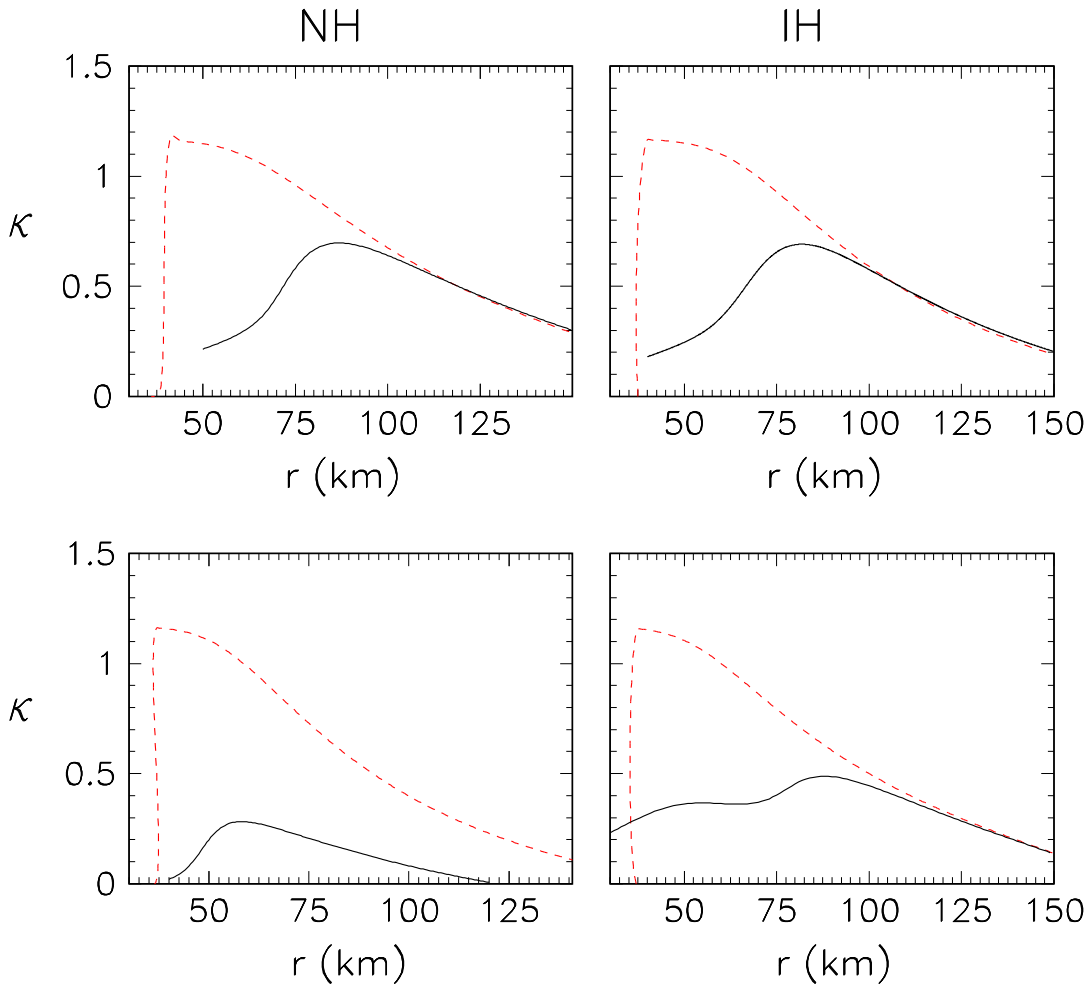} 
\caption{Case ${\cal C}$. Radial evolution of the $\kappa$ function in NH (left panel) and 
IH (right panel). Upper panels correspond to the bimodal instability, while the lower ones
to the MAA instability.  Continuous curves correspond to the MZA case while dashed ones to the SZA case.
} \label{fig6}
\end{figure*}

\begin{figure*}[!t]
 \includegraphics[angle=0,width=0.8\textwidth]{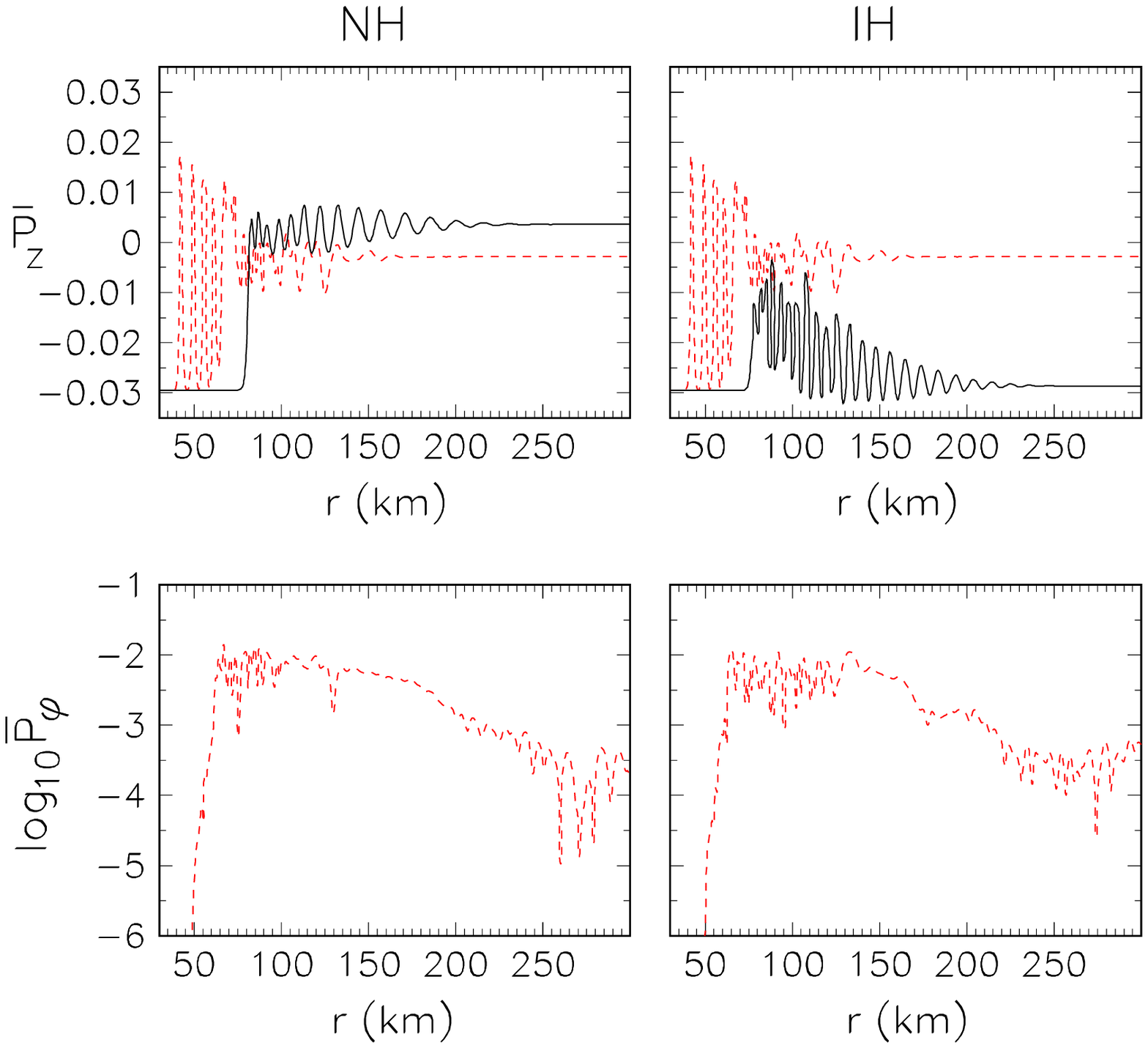} 
\caption{Case ${\cal C}$. Upper panels: Radial evolution of the integrated z-component 
${\bar P}_z$ of the polarization vector  for $\bar\nu$. 
Lower panels: Radial evolution of the integrated dipole
${\bar P}_\varphi$  for $\bar\nu$. 
Left panels refer to normal hierarchy, right panels to inverted hierarchy.
Dashed curves are for the MAA, SZA evolution; while
continuous curves are for the MAA, MZA case.
} \label{fig7}
\end{figure*}

The evolution of the $g_{\omega}$ is shown in Fig.~8 with the same format of Fig.~4. 
We realize that in the SZA case the MAA effects would produce  flavor decoherence with
a swap function $s(\omega)$ tending to zero. This is consistent with what found in~\cite{Mirizzi:2013rla} for the 
case of small flavor asymmetry. 
Conversely in the MZA case, since MAA effects are suppressed one recovers the splitting configurations already 
found in the axial symmetric case~\cite{Dasgupta:2009mg,Mirizzi:2010uz}.

The oscillated spectra for $e$ and $x$ flavor are shown in Fig.~9. 
These are very similar to what shown in~\cite{Dasgupta:2009mg,Mirizzi:2010uz} to which we address the reader for a detailed discussion. 
 Here we only remark that MZA effects  produce a smearing of the splitting features
with respect to what found in the SZA case.

\begin{figure*}[!t]
 \includegraphics[angle=0,width=0.8\textwidth]{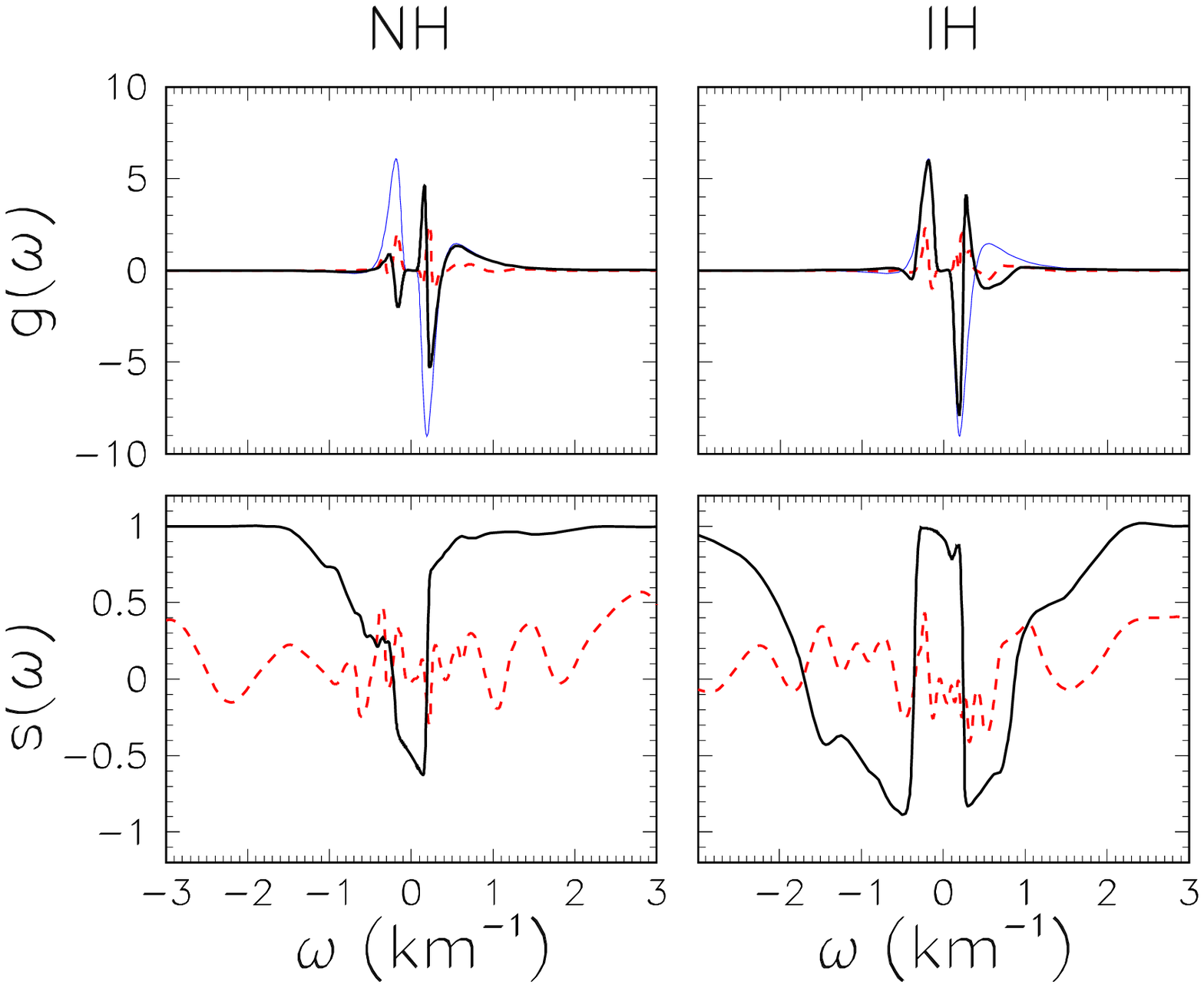} 
\caption{Case ${\cal C}$.  Upper panels: initial  (thin continuous curves) and final $g(\omega)$
for  the MAA evolution
in SZA (dashed curves) and MZA (thick continuous curves) case
 for normal hierarchy (left panel) and inverted hierarchy (right panel).
Lower panels: swap function, i.e. ratio of final with initial spectra 
in SZA  (dashed curves) and MZA (thick continuous curves).
} \label{fig8}
\end{figure*}

\begin{figure*}[!t]
 \includegraphics[angle=0,width=0.8\textwidth]{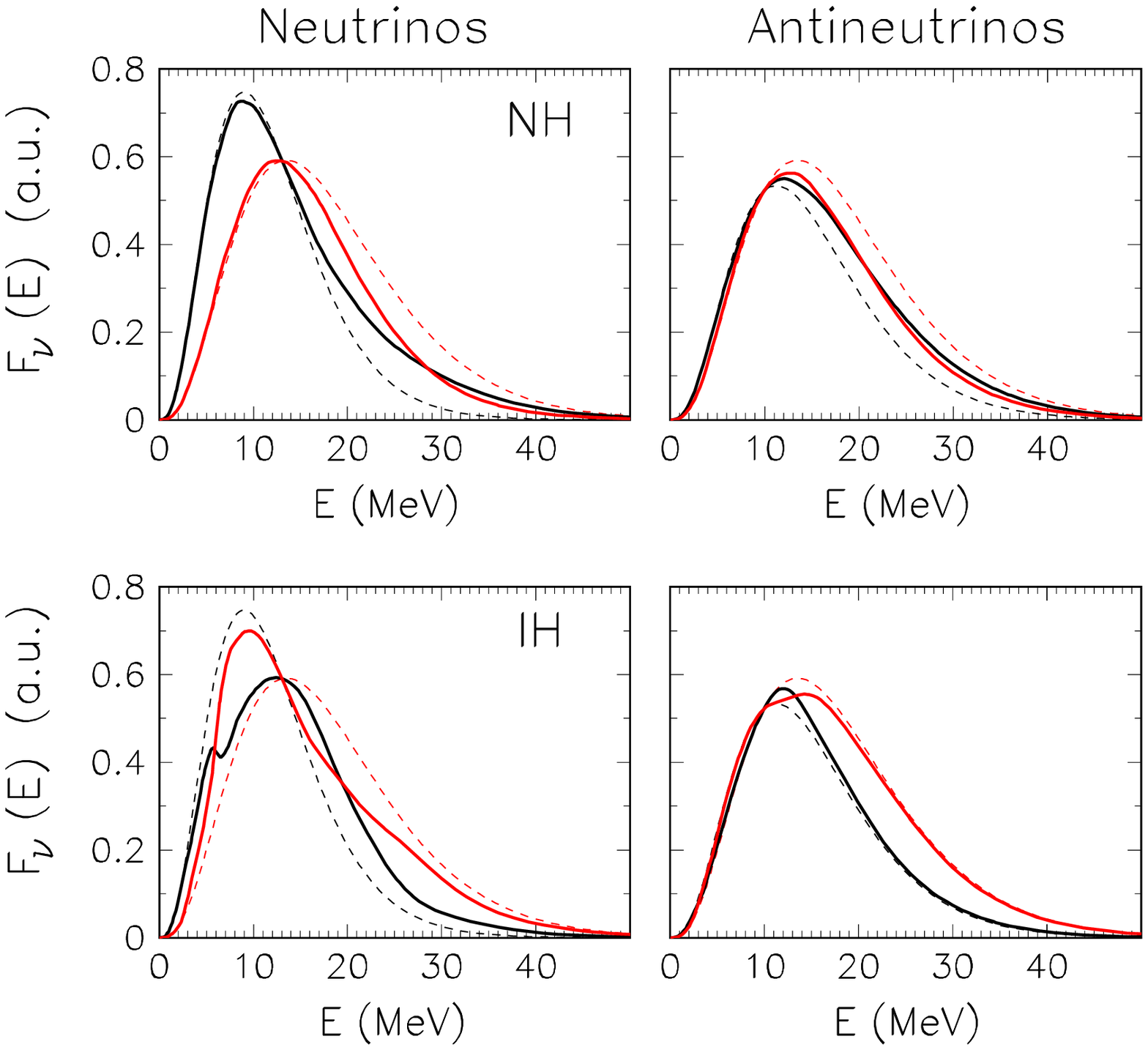} 
\caption{Case ${\cal C}$. MZA and MAA flavor evolution for $\nu$'s
(left panel) and $\bar\nu$'s (right panel) in NH (upper panels) and IH (lower panels).
 Energy spectra initially for
$\nu_e$ (black dashed curves) and $\nu_x$ (light dashed curves)
and after collective oscillations for $\nu_e$ (black continuous curves)
and $\nu_x$ (light continuous curves).
} \label{fig9}
\end{figure*}

\section{Conclusions}

It has been recently pointed out that removing the axial symmetry in the SN self-induced neutrino 
flavor evolution, MAA effects can trigger new flavor conversions. 
In this paper we have investigated the dependence of the MAA effects on the initial supernova neutrino fluxes.
We performed a stability analysis of the linearized equations of motion to classify the cases unstable under
the MAA effects. 
Then, we looked for a local solution along a given line of sight, working under the assumption that
the transverse variations of global solution are small.
If one considers  neutrino
fluxes typical of the accretion phase, with a pronounced
$\nu_e$ excess, \emph{de facto} behaving like spectra with a single
crossing, the MAA effects produce an instability in NH.
In our simplified scheme for the flavor evolution, this seems to produce spectral swaps and splits
very similar to what produced by the bimodal instability in IH, as expected from the 
recent analysis of~\cite{Raffelt:2013isa}. 

Another interesting spectral class is the one with neutrino fluxes having a moderate
flavor hierarchy and a $\nu_x$ excess, as possible during
the supernova cooling phase. In this situation, the
presence of multiple crossing points in the original
neutrino spectra destabilizes the synchronization at large
neutrino densities~\cite{Duan:2010bf,Mirizzi:2010uz}. In absence of synchronization,
in the SZA scheme collective conversions would
be possible at low-radii. 
In this case MAA effects would produce flavor decoherence in both the hierarchies.
 However, MZA effects suppress these low-radii conversions. We find that
 also MAA effects are suppressed and
one recovers the same splitting features expected in the axial symmetric case.

From our analysis it seems the  MAA effects play a crucial role only for spectra with a 
large flavor asymmetry. These would be natural during the accretion phase. 
However, it is known that during this  phase 
the  dense ordinary matter can be  large enough to potentially  inhibit the self-induced effects~\cite{Chakraborty:2011nf,Chakraborty:2011gd,Saviano:2012yh,Sarikas:2011am}.
In a recent work, this aspect has been investigated performing the stability 
analysis  with a realistic SN matter profiles. It has been found that during the accretion phase
the large matter term would suppress the MAA instability~\cite{Chakraborty:2014nma}.  
This result is  of crucial importance given the relevance of the supernova neutrino signal during the accretion phase to calibrate
the SN models~\cite{Tamborra:2013laa} and hopefully to extract the neutrino mass 
hierarchy~\cite{Serpico:2011ir,Borriello:2012zc}.

An interesting case where the MAA instability could play an important role is represented by the prompt $\nu_e$
deleptonization burst in low-mass O-Ne-Mg SNe. In this situation  the matter density profile can be so steep that the usual 
Mikheev-Smirnov-Wolfenstein (MSW) matter effects~\cite{Matt}
occur within the dense-neutrino region close to the neutrino sphere, preparing the conditions for the self-induced 
spectral splits~\cite{Duan:2007sh,Dasgupta:2008cd,Cherry:2013mv}. 
It would be interesting to see how the current picture is modified removing the axial symmetry. 

As we already mentioned, an open problem in the characterization of the MAA instability is that when the axial symmetry is broken, the global solution
of the neutrino equations of motion cannot be spherically symmetric. 
Presumably the axial symmetry breaking is
different in different overall directions. As a result,
there would be some finite ``transverse cell'' size with a common dipole of the azimuthal distribution.
If flavor conversions start relatively far from the neutrinosphere, neutrinos move mostly outward.
Therefore, one would expect that  they do not communicate in the
transverse direction, except from a bit of overlap among neighboring cells
at some larger radius, depending on the angular divergence of the neutrino
radiation. 
How large might be different transverse cells  is still an open issue. 
It  would probably depend on the initial
seeds, due to non-spherical neutrino emission
at the neutrinosphere.
These  may not vary fast as a function of direction because it averages over many
emission patches at the neutron star surface. Therefore,  as a function of observer direction,
the global solution presumably does  not change fast.
Under this assumption  the local solution along a given line of sight could give
an idea of what happens in the realistic situation. 
It is  mandatory to  test  if this working assumption is
correct, and if spectral features found in our simplified scheme are present also in the asymmetric
global solution. This task would be necessary in order to work out possible observable signatures
of the MAA instability.
 Getting the global solution would require
to  consider also variations along the  transverse
direction. This would imply passing from an ordinary to a partial differential equation problem, adding a big
layer of complication in the solution of the equations of motion. 

In conclusions, different gaps have still to be filled in order to achieve a better description
of the self-induced neutrino flavor conversions in SNe and  on the role of the multi-azimuthal-angle instability.
This effort is well motivated by the perspective of getting a reliable characterization of the SN neutrino
spectral features that would be observable in large underground detectors~\cite{Choubey:2010up}.

\section*{Acknowledgements} 

A.M. kindly thanks Georg Raffelt and Irene Tamborra for useful comments on the manuscript.
S.C.\ acknowledges support from the European Union through a Marie Curie Fellowship, Grant No.\ PIIF-GA-2011-299861, and through the ITN ``Invisibles'', Grant No.\ PITN-GA-2011-289442.
The work of A.M.  was supported by the German Science Foundation (DFG)
within the Collaborative Research Center 676 ``Particles, Strings and the
Early Universe.''


\end{document}